\documentclass[twocolumn,showpacs,pra,superscriptaddress]{revtex4}
\usepackage{dcolumn}
\usepackage{graphicx}
\usepackage{bbm}

\newcommand{\be}{\begin{equation}}
\newcommand{\ee}{\end{equation}}
\newcommand{\beq}{\begin{eqnarray}}
\newcommand{\eeq}{\end{eqnarray}}

\newcommand{\hide}[1]{}

\newcommand{\eq}[1]{Eq.\,(\ref{#1})}

\newcommand{\noeq}[1]{(\ref{#1})}
\newcommand{\fig}[1]{Fig.\,\ref{#1}}

\newcommand{\ket}[1]{\left| #1 \right\rangle}

\begin{document}

\title{Formation of deeply bound molecules via chainwise adiabatic passage}

\author{Elena Kuznetsova}
\affiliation{Department of Physics, University of Connecticut,
Storrs, CT 06269}
\affiliation{ITAMP, Harvard-Smithsonian Center
for Astrophysics, Cambridge, MA 02138} 
\author{Philippe Pellegrini}
\affiliation{Department of Physics, University of Connecticut,
Storrs, CT 06269}
\author{Robin C\^ot\'e}
\affiliation{Department of Physics, University of Connecticut,
Storrs, CT 06269} 
\author{M. D. Lukin}
\affiliation{Physics Department, Harvard University, Oxford St., Cambridge, MA 02138}
\author{S. F. Yelin}
\affiliation{Department of Physics, University of Connecticut,
Storrs, CT 06269} 
\affiliation{ITAMP, Harvard-Smithsonian Center
for Astrophysics, Cambridge, MA 02138}
\date{\today}

\begin{abstract}

We suggest and analyze a novel technique for efficient and robust creation of dense ultracold molecular ensembles in their ground rovibrational 
state. In our approach a molecule is brought to the ground state through a series of intermediate vibrational states via a {\em multistate 
chainwise Stimulated Raman Adiabatic Passage} (c-STIRAP) technique. We study the influence of the intermediate states decay on the transfer 
process and suggest an approach that minimizes the population of these states, 
 resulting in a maximal transfer efficiency. As an example, we analyze the formation of $^{87}$Rb$_{2}$ starting from an initial Feshbach molecular state and 
taking into account major decay mechanisms 
due to inelastic atom-molecule and molecule-molecule collisions. Numerical analysis suggests a transfer efficiency $>$ 90\%, even in the 
presence of strong collisional relaxation as are present in a high density atomic gas.

\end{abstract}

\maketitle

Ultracold molecular gases open possibilities for studyng new exciting physical phenomena and their applications. For example, ultracold molecules can find use in testing 
fundamental symmetries \cite{DeMille-EDM,Hudson}, in precision spectroscopy \cite{weak-inter,fine-structure} and ultracold chemistry \cite{quant-chem}. 
Dipolar ultracold quantum gases promise to show 
new phenomena due to strong anisotropic 
dipole-dipole interactions. Dipolar molecules in optical lattices can be employed as quantum simulators of condensed matter systems \cite{mol-cond-matt}. 
Ultracold polar molecules also represent an attractive platform for quantum computation \cite{DeMille}.

Dense samples of molecules in their ground rovibrational state $v=0,J=0$ are required for many of these applications. In this state, they have a large 
permanent electric dipole moment and are stable with respect to collisions and spontaneous emission. Currently translationally ultracold (100 nK - 1 mK) 
molecules are produced by 
magneto- \cite{Feshbach} and photoassociation \cite{photoass} techniques. In both of these techniques the 
molecules are translationally cold, but vibrationally hot, since they are formed in high vibrational states near the dissociation limit of the electronic ground state. 
Therefore, once created, molecules have to be rapidly transfered 
to the ground rovibrational state. 

One of the most efficient ways to transfer population between two states is based on the Stimulated Raman Adiabatic Passage (STIRAP) technique \cite{STIRAP-review, 
STIRAP-experiment,STIRAP-KRb,STIRAP-Ye}. 
STIRAP provides a lossless robust transfer between an initial and a final state of a three-level system using a Raman 
transition with two counterintuitively ordered laser pulses. The 
main difficulty with a two-pulse STIRAP in molecules is to find an intermediate vibrational state of the excited electronic potential with a good Franck-Condon overlap
 with both 
a highly delocalized initial high vibrational state and a tightly localized $v=0$ state \cite{Stwalley}. It was therefore 
proposed in \cite{Zoller} to transfer population in several steps 
down the ladder of vibrational states using a sequence of stimulated optical Raman transitions. In this case the initial and final 
vibrational levels of each step do not differ significantly, and it is easier to find a suitable intermediate vibrational level in the excited electronic state. 
In this step-wise approach population 
is transferred through a number of vibrational levels in the ground electronic state. In a dense gas, molecules in such states are subject to inelastic collisions with 
 background atoms or other molecules. The released kinetic energy greatly exceeds the trap depth resulting in loss of both molecules and atoms
from the trap. This process is expected to limit the efficiency of creation of dense ultracold molecular samples. 
In this work we present a technique allowing an efficient transfer of a molecule from a high-lying to the ground vibrational state 
which minimizes population loss due to inelastic collisions in intermediate levels. Our technique is based on generalized chainwise STIRAP, which in 
principle allows for lossless transfer to the ground vibrational state. We note that serial STIRAP as in \cite{Zoller,STIRAP-Ye} and pump-dump technique with a train of 
short pulses \cite{Jun-pump-dump} should also allow lossless transfer if pulses are shorter than the collisional relaxation time. 

\begin{figure}
\center{
\includegraphics[width=6.5cm]{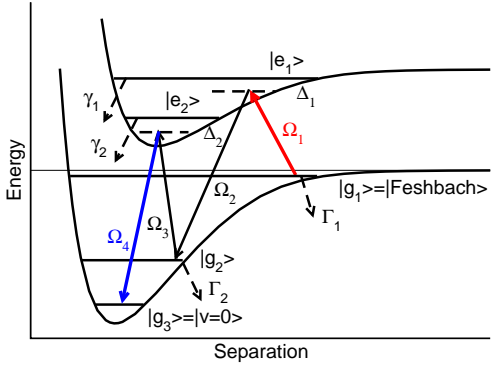}
\caption{\label{fig:chain-STIRAP}Schematic showing the multistate chainwise STIRAP transfer of population from the Feshbach $\ket{g_{1}}$ to the ground $\ket{g_{3}}$ 
vibrational state.}
}
\end{figure}

The idea of this work can be described using a simple five-level model molecular system with states chainwise coupled by optical fields as illustrated in \fig{fig:chain-STIRAP}.  
The states $\ket{g_{1}}$, $\ket{g_{2}}$ and $\ket{g_{3}}$ are vibrational levels of the ground electronic molecular state, while $\ket{e_{1}}$ and $\ket{e_{2}}$ are 
vibrational 
states of an excited electronic molecular state. Molecules are formed in a high vibrational state $\ket{g_{1}}$, which in the following ia assumed to be a molecular 
Feshbach state. The state $\ket{g_{3}}$ is the deepest bound vibrational state $v=0$, and $\ket{g_{2}}$ is an intermediate 
vibrational state. The goal is to efficiently 
transfer population from the state $\ket{g_{1}}$ to state $\ket{g_{3}}$. At least two vibrational 
levels $\ket{e_1}$ and $\ket{e_2}$ in an excited electronic state are 
required, one having a good Franck-Condon overlap with $\ket{g_3}$, and the other with the initial Feshbach molecular state $\ket{g_1}$. 

In the states $\ket{e_{1}}$ and $\ket{e_{2}}$, molecules decay due to spontaneous 
emission and collisions, 
and in the states $\ket{g_{1}}$ (for bosonic molecules) and $\ket{g_{2}}$ they experience 
fast inelastic collisions with background atoms leading to loss of molecules from a trap. It means that populating the states $\ket{e_{1}}$, $\ket{e_{2}}$ and 
$\ket{g_{2}}$ has to be avoided when a background atomic gas is present, or the transfer process has to be faster than the collisional relaxation time.

This can be achieved via chainwise STIRAP. The wave function of the system is $\ket{\Psi}=\sum_{i}C_{i}\exp{(-i\phi_{i}(t))}\ket{i}$, where 
$i=g_{1},e_{1},g_{2},e_{2},g_{3}$; $\phi_{g_{1}}=0$, $\phi_{e_{1}}=\nu_1t$, 
$\phi_{g_2}=(\nu_2-\nu_1)t$, 
$\phi_{e_{2}}=(\nu_3+\nu_2-\nu_1)t$, $\phi_{g_{3}}=(\nu_4-\nu_3+\nu_2-\nu_1)t$; $\nu_i$ is the frequency of the $i$th 
optical field. The evolution is then governed by the 
Schr\"odinger equation
\begin{equation}
i\hbar \frac{\partial \ket{\Psi}}{\partial t}=H(t)\ket{\Psi},
\end{equation}
where the time-dependent Hamiltonian is given by 
\begin{equation}
\label{eq:hamiltonian}
H = \left( \begin{array}{ccccc}
0 & -\Omega_{4} & 0 & 0 & 0 \\
-\Omega_{4} & \Delta_{2} & -\Omega_{3} & 0 & 0 \\
0 & -\Omega_{3} & 0 & -\Omega_{2} & 0 \\
0 & 0 & -\Omega_{2} & \Delta_{1} & -\Omega_{1} \\
0 & 0 & 0 & -\Omega_{1} & 0 \end{array} \right).
\end{equation}
Here 
$\Omega_{i}(t)=\mu_i{\cal E}_{i}(t)/2\hbar$, $i=1,2,3,4$ are the Rabi frequencies of optical fields; ${\cal E}_{i}$ is the amplitude of $i$th optical field, 
$\mu_{i}$ is the dipole matrix element along the respective transition,  $\Delta_{1}=\omega_{1}-\nu_{1}$ and $\Delta_{2}=\omega_{4}-\nu_{4}$ are 
one-photon detunings of the fields, and the $\omega_i$ are the molecular frequencies along transition $i$. 
We assumed in \eq{eq:hamiltonian} that pairs of fields coupling two neighboring ground state vibrational levels are in a two-photon (Raman) 
resonance. 

The Hamiltonian \eq{eq:hamiltonian} has a dark state, a specific superposition of states uncoupled from applied laser fields, given by the expression 
\begin{equation}
\label{eq:dark-state}
\ket{\Phi^{0}}=\frac{\Omega_{2}\Omega_{4}\ket{g_{1}}-\Omega_{4}\Omega_{1}\ket{g_{2}}+\Omega_{1}\Omega_{3}\ket{g_{3}}}{\sqrt{\Omega_{4}^{2}\Omega_{1}^{2}+\Omega_{1}^{2}\Omega_{3}^{2}+\Omega_{2}^{2}\Omega_{4}^{2}}}.
\end{equation}
In c-STIRAP (as in classical STIRAP) the optical fields are applied in a counterintuitive way, i.e. at $t=-\infty$ only a combination of the $\Omega_{4}$, $\Omega_{3}$, $\Omega_{2}$ fields, and at 
$t=+\infty$ only of $\Omega_{3}$, $\Omega_{2}$, and $\Omega_{1}$ is present. As a result the dark state is initially 
associated with the $\ket{g_{1}}$ and finally with the $\ket{g_{3}}$ state. Adiabatically changing the Rabi frequencies of the optical fields 
so that the system stays in the dark state during evolution, one can 
transfer the system from the initial high-lying $\ket{g_{1}}$ to the ground vibrational $\ket{g_{3}}$ state with unit efficiency, defined as the population of the $\ket{g_{3}}$ 
state at $t=+\infty$.
The dark state does not have contributions from the $\ket{e_{1}}$ and $\ket{e_{2}}$ excited
states, which means that they are not populated during the transfer process. As a result, the decay from these states does not 
affect the transfer efficiency. To minimize the population in the intermediate ground vibrationl state we apply the fields such that 
$\Omega_{2}$, $\Omega_{3}\gg$ $\Omega_{1}$, $\Omega_{4}$ and $\Omega_{2}$, 
$\Omega_{3}$ temporally overlap both the $\Omega_{1}$ and $\Omega_{4}$ pulses \cite{Malinovsky}. In this case 
Eq.(\ref{eq:dark-state}) indicates that the population in the $\ket{g_{2}}$ state can, in principle, be completely suppressed at all times. 
This is the main idea of this work.

We now analyze this system in detail. To simplify the analysis, we assume $\Omega_{2}=\Omega_{3}=\Omega_{0}$; $\Omega_{0}$ is independent of time (in practice the corresponding pulses just have to be 
much longer than $\Omega_{1}(t)$, $\Omega_{4}(t)$ and overlap both of them), and $\Omega_{0}\gg |\Omega_{1}|,|\Omega_{4}|$. We also set one-photon detunings to zero $\Delta_{1}=\Delta_{2}=0$, and 
define the effective Rabi frequency $\Omega(t)=\sqrt{\Omega_{1}^{2}+\Omega_{4}^{2}}$ and a rotation angle by $\tan\theta(t)=\Omega_{1}(t)/\Omega_{4}(t)$. 
The 
eigenvalues of the system 
(2) are $\varepsilon_{0}=0$, corresponding to the dark state,  and $\varepsilon_{1,2}=\pm \Omega/\sqrt{2}$, and 
$\varepsilon_{3,4}=\pm \sqrt{2}\Omega_{0}$ to bright states. Adiabatic eigenstates $\ket{\Phi}=\left\{\ket{\Phi^{n}}\right\}$, $n=0,...4$ and 
 the bare states are transformed as $\Psi_{i}=\sum_{j}W_{ij}\Phi_{j}$
via a rotation matrix 

\begin{equation}
W = \frac{1}{2} \left( \begin{array}{ccccc}
-2\cos\theta & 0 & \xi \sin2\theta & 0 & -2\sin\theta \\
-\sqrt{2}\sin\theta & 2 & -\frac{\xi}{\sqrt{2}}\cos2\theta & -1 & \sqrt{2}\cos\theta \\
-\sqrt{2}\sin\theta & -1 & -\frac{\xi}{\sqrt{2}}\cos2\theta & 1 & \sqrt{2}\cos\theta \\
\xi\sin\theta & -1 & \sqrt{2} & -1 & \xi\cos\theta \\
\xi\sin\theta & 1 & \sqrt{2} & 1 & \xi\cos\theta \end{array} \right),
\label{(A2)}
\end{equation}
where $\xi=\Omega/\Omega_{0}$ and terms of the order of $O(\xi^{2})$ and higher are neglected.

The adiabaticity condition in this case requires $\dot{\theta}\ll \Omega,\;\Omega_{0}$. If the condition $\dot{\theta}\ll \Omega$ is satisfied, the dark state will not couple to the 
$\ket{\Phi^{1,2}}$ states, corresponding to the closest in energy $\varepsilon_{1,2}$ eigenvalues. Coupling to the $\ket{\Phi^{3,4}}$ states will be suppressed even more strongly, since $\Omega \ll \Omega_{0}$.
This gives a standard STIRAP adiabaticity requirement $\Omega T_{tr}\gg 1$, where $T_{tr}$ is the c-STIRAP transfer time.

To study the effect of the decay from $\ket{g_{2}}$ and $\ket{g_{1}}$ on the dark state evolution, 
we turn to a density matrix description \cite{STIRAP-with-loss}, and use the adiabatic basis states. 
The density matrix equation then takes a form 
\begin{equation}
\label{eq:den-matr-a}
i\hbar \frac{d\rho^{a}}{dt}=\left[H^{a},\rho^{a}\right]-i\hbar\left[W^{T}\dot{W},\rho^{a}\right]-{\cal L}^a\rho^a,
\end{equation}
where the density matrix $\rho$ and the Liouville operator ${\cal L}$ in this basis are given by $\rho^{a}=W^{T}\rho W$ and ${\cal L}^a\rho^a=W^{T}{\cal L}\rho W$, and the 
Hamiltonian $H^{a}$ is diagonal; $W$ is the rotation matrix. The Liouville operator ${\cal L}$ consists of the usual decays, where only population decays ($\propto T_1^{-1}$) into other vibrational states or the continuum 
are considered (see Fig.\ref{fig:chain-STIRAP}). Since at $t=-\infty$ all population is assumed to be in state $\ket{g_1}$, initial conditions for 
\eq{eq:den-matr-a} read as $\rho^{a}_{00}=1$, $\rho^{a}_{nm}=0$ for $nm\neq 00$, where 
$\rho^{a}_{00}$ denotes the dark state population.

The decay of the dark state due to the population loss from the $\ket{g_{1}}$ and $\ket{g_{2}}$ states is then described by the equation 
(keeping only terms up to the $\Omega^{2}/\Omega_{0}^{2}$ order)

\begin{equation}
\label{eq:dark-state-chainwise}
\frac{\dot{\rho}^{a}_{00}}{\rho^{a}_{00}}\approx
 -(\Gamma_{2}+\Gamma_{1}\cos^{2}\theta)\left(\frac{\Omega}{2\Omega_{0}}\sin 2\theta\right)^{2}-
 \Gamma_{1} \cos^{2}\theta. \nonumber
\end{equation}
Equation \noeq{eq:dark-state-chainwise} shows that the intermediate state decay can be neglected during the transfer time $T_{tr}$ if $(\Gamma_{1}+\Gamma_{2})T_{tr}\left(\sin2\theta\Omega/2\Omega_{0}\right)^{2}\ll1$. 
From this expression one can see that the intermediate state decay rate is reduced by a factor $(\Omega/\Omega_{0})^{2}\ll1$ in this regime. It also follows from 
\eq{eq:dark-state-chainwise} that decay from $\ket{g_{1}}$ is not suppressed, so that the transfer process has to be faster than this decay.

Magneto- and photo-association techniques produce molecules mostly from ultracold Bose, two-spin component Fermi and mixture of alkali metal atomic gases. 
In traps with high initial atomic density, weakly bound Feshbach molecules rapidly decay due to inelastic atom-molecule collisions, which were found to be the 
major limiting factor of molecule lifetime. 
Depending on the quantum statistics of the constituent atoms, the alkali dimers show different behavior with respect to inelastic atom-molecule and molecule-molecule collisions. 
Fermionic alkali dimers in the Feshbach state are very stable with respect to collisions, especially close to the 
resonance, where the scattering length is large. Lifetimes of the Feshbach molecules of the 
order of 1 s have been observed experimentally \cite{Feshbach-lifetime,Feshbach-lifetime-1}.
In contrast, bosonic and mixed dimers experience fast vibrational quenching due to inelastic atom-molecule collisions, even in their Feshbach state. 
An atomic density in a trap is typically in the range 
$n_{at}\sim 10^{11}-10^{14}$ cm$^{-3}$, then the Feshbach state relaxation rate is in the range $\Gamma_{1}\sim 10^{1}-10^{4}$ s$^{-1}$ (calculated from the 
corresponding inelastic atom-molecule collision coefficient $k_{inel}\sim 10^{-10}$ cm$^{3}$s$^{-1}$ 
\cite{Ketterly,Cs,Rb-collisions}). At the same atomic densities the vibrational 
relaxation rate $\Gamma_{2}$ of intermediate vibrational states for bosonic molecules is in the range  
$\Gamma_{2}\sim 10^{2}-10^{5}$ s$^{-1}$ (calculated from $k_{inel}\sim 6\cdot 10^{-10}$ cm$^{3}$s$^{-1}$ for $^{7}$Li$_{2}$ \cite{collisions-review} 
and the same range of atomic densities). 
Inelastic molecule-molecule collisional relaxation rates are about two orders of magnitude smaller due to typically smaller molecular density.

\begin{figure}
\center{
\includegraphics[width=6.cm]{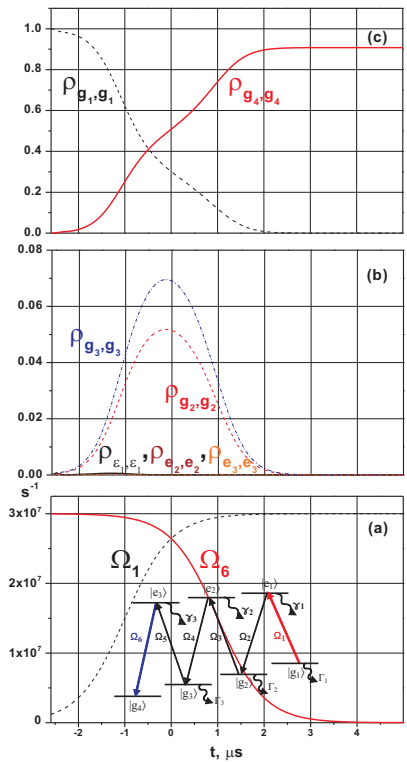}
\caption{\label{fig:STIRAP-b}Results of numerical solution of the density matrix equation for a seven-state $^{87}$Rb$_{2}$ molecular system: a) pump $\Omega_{1}$ (solid line) and 
Stokes $\Omega_{6}$ (dashed line) Rabi frequencies; b) populations of the intermediate vibrational states; c) populations of the Feshbach $\rho_{g_{1}g_{1}}$ (dashed line) 
and the ground $v=0$ state $\rho_{g_{3}g_{3}}$ (solid line). 
Parameters used: $\Gamma_{1}=10^{4}$ s$^{-1}$, 
$\Gamma_{2}=\Gamma_{3}=6\cdot 10^{4}$ s$^{-1}$, $\gamma_{1}=8\cdot 10^{7}$ s$^{-1}$, $\gamma_{2}=\gamma_{3}=3\cdot 10^{7}$ s$^{-1}$; $\Omega_{1}^{max}=\Omega_{6}^{max}=3\cdot 10^{7}$ s$^{-1}$, 
$\Omega_{i}=1.2\cdot D_{v,v'\;i} \cdot 10^{8}$ s$^{-1}$ i=2,3,4,5, $T=1$ $\mu$s, $\tau=-2$ $\mu$s for both schemes. Inset in (a) shows a model Rb$_{2}$ system with decays.
}
}
\end{figure}

We next illustrate the technique for a sample seven-state bosonic 
$^{87}$Rb$_{2}$ molecular system (see the inset to Fig.\ref{fig:STIRAP-b}a). In the first step, the Feshbach state can be 
coupled to the electronically excited pure long range molecular state $\ket{0_{g}^{-},v,J=0}$, located close to the $5S_{1/2}+5P_{3/2}$ 
dissociation asymptote. For example, following \cite{STIRAP-experiment}, the $v=31$ vibrational level can be chosen $6.87$ cm$^{-1}$ below the 
dissociation limit. The second STIRAP step can be to $v=116$ in the ground electronic state. The authors of Ref.\cite{STIRAP-experiment}
mention that the Franck-Condon factors from the excited $\ket{0_{g}^{-},v=31,J=0}$ state to the ground state vibrational levels down to the 
$X\;^{1}\Sigma^{+}_{g}(v=116)$ are similar to the second-to-last vibrational state used in the STIRAP experiment in \cite{STIRAP-experiment}.
The ground $v=0$ state can then be reached in four steps, using e.g. the path given in Table I. We note that in Rb$_{2}$ the $v=0$ state cannot be 
reached from the $v=116$ in two steps due to unfavorable Franck-Condon factors, a minimum of four steps is therefore 
required, resulting in a seven-state system. A four-step path from the Feshbach to the $v=0$ state can be realized in 
Cs$_{2}$ \cite{Cs-STIRAP}, and therefore in other alkali dimers as well.

\begin{table}
\centering
\caption{Possible chainwise transfer path from the Feshbach to the ground rovibrational state in the Rb$_{2}$ molecule. Shown also 
the corresponding transition dipole moments and wavelengths.}
\begin{tabular}{c c c c}
\hline
i& $v - v'$ transition & $D_{v,v'i}$  & $\lambda$ \\
&                    &       Debye    &     nm  \\
\hline
1 & $\ket{Feshbach} - \ket{0^{-}_{g},v=31,J=0}$ & 0.4 & 780.7 \\
2 & $\ket{0^{-}_{g},v=31,J=0} - X\;^{1}\Sigma^{+}_{g}(v=116,J=0)$ & 0.8 & 780.4 \\
3 & $X\;^{1}\Sigma^{+}_{g}(v=116,J=0) - A^{1}\Sigma^{+}_{u}(v'=152,J=1)$ & 0.55\ & 846 \\
4 & $A\;^{1}\Sigma^{+}_{u}(v'=152,J=1)- X^{1}\Sigma^{+}_{g}(v=50,J=0)$ & 0.64 & 907.4 \\
5 & $X\;^{1}\Sigma^{+}_{g}(v=50,J=0)- A^{1}\Sigma^{+}_{u}(v'=21,J=1)$ & 0.53 & 990 \\
6 & $A\;^{1}\Sigma^{+}_{u}(v'=21,J=1)- X^{1}\Sigma^{+}_{u}(v=0,J=0)$ & 2.37 & 856.4 \\
\hline
\end{tabular}
\label{table:Rb}
\end{table}
The transitions $\ket{e_{1}}-\ket{g_{2}}$; $\ket{g_{2}}-\ket{e_{2}}$, $\ket{e_{2}}-\ket{g_{3}}$ and 
$\ket{g_{3}}-\ket{e_{3}}$ are coupled by CW laser fields, the first transition $\ket{g_{1}}-\ket{e_{1}}$ and the last transition $\ket{e_{3}}-\ket{g_{4}}$ in the chain 
are coupled by the fields
$\Omega_{1}=\Omega_{1}^{max}(1+\tanh{(t-\tau/2)/T})/2$ and $\Omega_{6}=\Omega_{6}^{max}(1-\tanh{(t+\tau/2)/T})/2$, respectively. 
In the above scheme we picked the $\ket{e_{3}}-\ket{g_{4}}$ transition with a large transition dipole moment, and intermediate 
transitions coupled by CW fields with close and reasonably large moments. In this case the Stokes pulse intensity can be minimized, 
and CW fields, provided, e.g., by laser diodes, can have the same intensity to optimize the transfer efficiency. The wavelengths of the transitions in 
Table I are covered by Ti:Sapphire and diode lasers. To provide the phase coherence between the laser fields required to carry out STIRAP, lasers can be 
phase locked to spectral components of a frequency comb \cite{freq-comb}.

The results of the numerical simulation are given in \fig{fig:STIRAP-b}. 
We assumed that the CW fields have 
the same amplitude of the electric field ${\cal E}_{0}$, resulting in a Rabi frequency $\Omega_{0\;i}=D_{v,v'i}{\cal E}_{0}/2\hbar$ for $i$th transition.
The Rabi frequencies of STIRAP 
fields were chosen to satisfy a condition that $\Omega$ is less than the binding energy of the Feshbach molecular state to minimize Raman dissociation of weakly bound molecules. The pulse duration $T$ and delay $\tau$ 
were varied to obtain the maximal transfer efficiency. To estimate the decay rate of intermediate vibrational states, the highest atomic density $n_{at}\sim 10^{14}$ cm$^{-3}$ 
available experimentally was used along with the inelastic collision coefficient for intermediate vibrational states $k_{inel}\sim 6\cdot 10^{-10}$ cm$^{3}$s$^{-1}$, giving $\Gamma_{2,3}=6\cdot 10^{4}$ s$^{-1}$. 
A Decay rate of the Feshbach state $\Gamma_{1}=10^{4}$ s$^{-1}$ was used. Numerical analysis shows that $>90\%$ of the population can be 
transferred to $v=0$ at high initial atomic density even in the presence of collisional decay from the intial Feshbach state. As can be seen from 
Fig.\ref{fig:STIRAP-b}b, the population of the 
intermediate ground vibrational states does not exceed $7\%$ during the transfer process and only for a short time, reducing the molecular loss due to collisions in this states.

We can now estimate intensities of CW and pulsed fields corresponding to Rabi frequencies used in our calculations. Taking the peak Rabi 
frequency of the pump and Stokes fields $\Omega_{1,6}^{max}=3\cdot 10^{7}$ s$^{-1}$, the corresponding intensities are
$I_{1,6}^{peak}=c{\cal E}_{1,6}^{2}/8\pi=c(\Omega_{1,6}^{max}\hbar/D_{v,v'\;1,6})^{2}/8\pi$, resulting in $I_{1}\sim 3$ W/cm$^{2}$ and 
$I_{6}\sim 0.1$ W/cm$^{2}$; 
for CW fields with a Rabi frequency $\Omega_{0\;i}\sim 6\cdot 10^{7}$ s$^{-1}$ the corresponding intensity is $I_{2,3,4,5}\sim 5$ W/cm$^{2}$.

In summary, we propose a method of vibrational cooling of ultracold molecules, based on the multistate 
chainwise STIRAP technique. Molecules which are formed in high-lying vibrational states are 
transfered into a ground rovibrational state $v=0,J=0$ using Raman transitions via several intermediate vibrational states in 
the ground electronic state. Our technique provides 100$\%$ vibrational as well as rotational selectivity using selection rules 
$\Delta J=0,\pm 1$ for rotational transitions.
Numerical analysis of the transfer process for a typical bosonic Rb$_{2}$ molecular system in a trap with a high atomic density $n_{at}\sim 10^{14}$ cm$^{-3}$ shows that 
transfer efficiencies $\sim 90\%$ are 
possible even in the presence of fast collisional relaxation of the Feshbach molecular state. 

The multistate chainwise STIRAP technique allows one to use various transitions, coupled by, e.g., rf fields and DC interactions. It can therefore be 
combined with 
the recently demonstrated resonant association method \cite{Res-assos}. Another possibility is to use the magnetic field dependent DC interchannel coupling between an 
entrance and a closed channel state as a first transition in the STIRAP chain \cite{STIRAP-DC} followed by optical transitions to the ground vibrational state.
The chainwise STIRAP can be applied to resonant photoassociation as 
well, then the first transition in the STIRAP chain will couple the continuum states to a high energy vibrational state in the ground electronic state \cite{Robin}.

We gratefully acknowledge fruitful discussions with J. Ye and financial support from ARO and NSF.

\end{document}